# High gas pressure and high-temperature synthesis (HP-HTS) technique and its impact on iron-based superconductors

Mohammad Azam[1], Manasa Manasa[1], Andrzej Morawski[1], Tomasz Cetner[1], and Shiv J. Singh[1,*]

[1] Institute of High Pressure Physics (IHPP), Polish Academy of Sciences, Sokołowska 29/37, 01-142 Warsaw, Poland

* Correspondence: sjs@unipress.waw.pl

**Abstract:** The high-pressure growth technique generally plays an important role in the improvement of the sample quality and the enhancement of various physical and magnetic properties of materials. The high gas pressure technique provides a large sample space (10-15 cm) to grow various kinds of materials. In this paper, we introduce the high gas pressure and high-temperature synthesis (HP-HTS) technique that is present at our institute and is applied to the growth process of different kinds of superconducting materials, particularly iron-based superconductors. More details and the working principle of this HP-HTS technique are discussed. We have also demonstrated the current results based on the iron-based superconductors by using this unique HP-HTS technique. These results demonstrate the enhancement of the superconducting properties with the improved sample quality compared to the conventional synthesis process at ambient pressure.

## 1. Introduction

High-pressure synthesis of the material is an important area for physics, chemistry, and material sciences [1, 2, 3]. This method reduces the chemical reaction time, controls the evaporation of lighter elements [4], and can also be used to grow new materials that cannot be prepared at ambient pressure, such as superhard materials like diamond and cubic boron nitride [5]. Generally, the sample space is very tiny for the pressure growth process, and due to this, a small sample size is always obtained [6]. This issue has been clearly observed with the growth process of iron-based superconductor (FBS) [7]. To overcome this problem, we need to find a good high-pressure growth method that can provide large crystals and large amount of bulk samples with high superconducting properties. However, the question is: which technique is more suitable to resolve these problems? [8] [9].

There are two kinds of pressure techniques: *a) Solid-medium pressure techniques* such as Hot Isostatic Pressure (HIP), Diamond Anvil Cell (DAC) Technique, Multi-Anvil High-Pressure Apparatus, and cold synthesis [10]. The properties of this technique are as follows: (i) It has a limited sample space of up to 0.5 cm$^3$, and (ii) the pressure and temperature distribution are not homogeneous. It generates undefined preparation conditions. (iii) Due to the pressure medium touching the sample, there is a great possibility of introducing the impurity phases. (iv) Also, the temperature gradient is not easy to control in the multizone furnace. (v) And always, the sample size is smaller due to the small sample space. *b) Gas pressure technique*: The properties are as follows: (i) it has several cm$^3$ of sample space, and (ii) it's easy to create homogeneous temperature and pressure stability for practically long growth time. (iii) The spatial temperature profile may be controlled and it may be easy to control the partial gas pressure; (iv) Growth of the crystal is a comparatively easy process; (v) No possibility of introducing impurities from the pressure medium; (vi) Sample chamber with an internal three-zone furnace with pressure up to 2-3 GPa and temperature ~2000°C.

In solid-medium pressure growth process, interactions between materials and instrument parts can introduce contamination and other issues due to the applied physical force, potentially leading to morphology problems like cracks and pores [1]. Therefore, it is crucial to explore alternative techniques to address these challenges. These comparative studies suggest that the gas pressure technique can be a unique and attractive way to grow high-quality and large amounts of samples and to improve the superconducting properties of FBS [11] [12]. These reasons motivated us to use high-pressure techniques for the growth of iron-based superconductors.

In the case of FBS, few studies have been reported based on solid-pressure medium techniques where the sample size is enhanced from 10 micrometers to 300 micrometers with improved superconducting properties [6, 7]. These reports suggest that more studies are needed in this direction by using different pressure techniques, such as the high-gas growth method so that we can also prepare a large amount or large size sample with high superconducting properties. In this paper, we have introduced the principle and more details about our high gas pressure and high-temperature synthesis (HP-HTS) technique that is available at our institute, 'UNIPRESS'. Also, the current results based on iron-based superconductors using this HP-HTS method will be presented and discussed.

## 2. Principle of high-pressure technique

This HP-HTS has a compressor, called the reciprocating compressor which is based on the oil gas piston. The main working principle of the reciprocating compressor is based on Boyel's law [13] which states that absolute pressure exerted by a given mass of an ideal gas is inversely proportional to the occupied volume. The piston compresses the gas and increases the gas pressure inside the chamber. Generally in our HP-HTS system, there are three piston cavities (piston chamber) attached to each other in a series manner to achieve the required high pressure. The systematic block diagram of the working principle of these pistons is shown below in various stages:

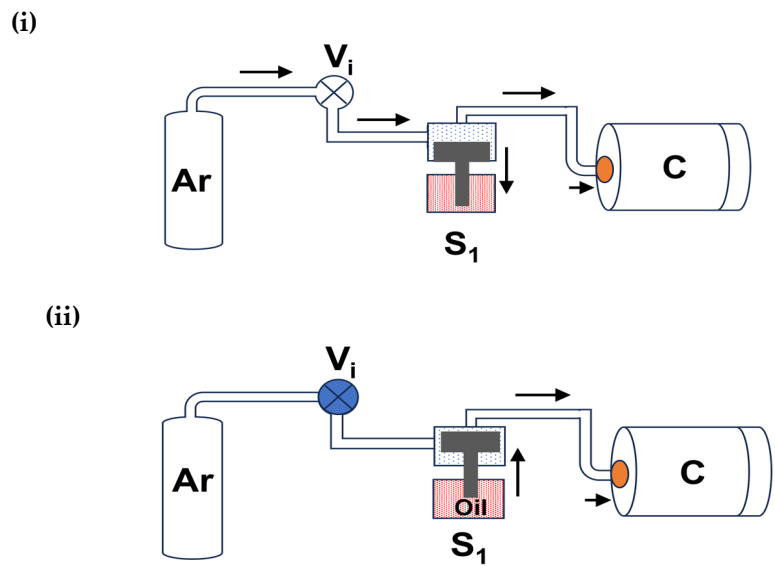

**Figure 1:** The block diagram of the first-stage gas compression process: **(i)** Gas flow into the $S_1$ cavity and chamber (C); **(ii)** $S_1$ piston compressing the gas into chamber (C) up to 800 bar.

*(a) First-stage piston ($S_1$):* First, the gas bottle is opened, and the gas enters through the intake valve ($V_i$) into all piston cavities and the high-pressure chamber. Figure 1(i) demonstrates the gas refilling process in the first piston cavity (S$_1$) and chamber (C). During this process, gas is filled in the upper part of the S$_1$ by moving the piston down to the oil tank. Basically, the piston moves down with the gas bottle pressure of ~200 bar. After 4-5 minutes, we close the intake valve (V$_i$) and move the piston in an upward direction slowly to pressurize the gas inside the chamber. Figure 1(ii) depicts the movement of the piston in the upward direction in the S$_1$ cavity. Through this first stage, we can reach the maximum pressure up to 800 bar, *i.e.*, the gas pressure can increase from 200 bar to 800 bar.

*(b) Second-stage piston (S$_2$)*: This stage is connected to the first-stage piston (S$_1$), and has a pressure of around 800 bar. There is again an oil-gas piston that moves down with the first-stage pressure of 800 bar, which is shown in Figure 2(i). Now, the intake valve (V$_i$) between S$_1$ and S$_2$ is closed, and the piston of S$_2$ moves upward slowly which creates gas pressure inside the chamber up to 4000 bar. This process is similar to the first stage and is shown in Figure 2 (ii).

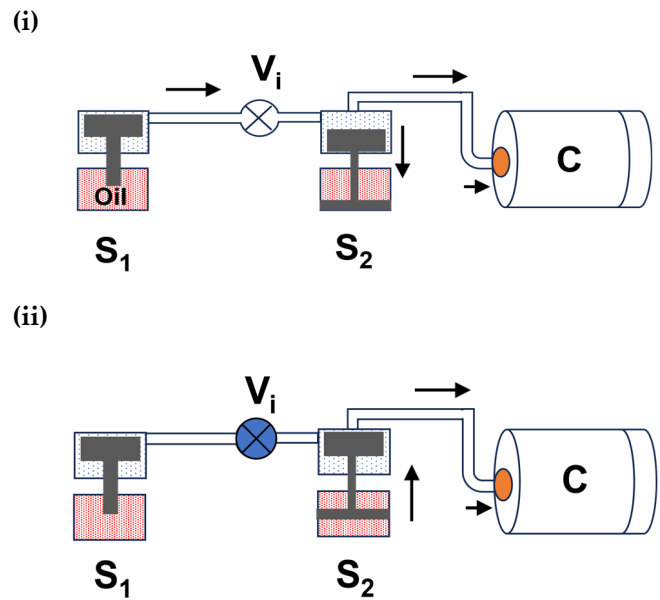

**Figure: 2.** The block diagram of the second stage gas compression process: **(i)** Gas flow into the S$_2$ cavity and chamber (C) **(ii)** S$_2$ piston compressing the gas into chamber (C) up to 4000 bars.

*(c) Third stage piston (S$_3$):* This stage piston has bigger size than the first and second-stage pistons. Due to the second-stage pressure, this piston is moved to the down position, which is depicted in Figure 3(i). Once all gas pressure moves into the S$_3$ cavity and chamber, the intake valve (V$_i$) between S$_2$ and S$_3$ is closed. In the next step, S$_3$ piston move slowly upward direction through the oil-based pump and enhances the pressure inside the chamber (C). The maximum pressure achieved in S$_3$ can be reached up to 1.8 GPa (18000 bar). Figure 3(ii) displays the block diagram of the third stage piston movement to achieve the highest pressure.

**(i)**

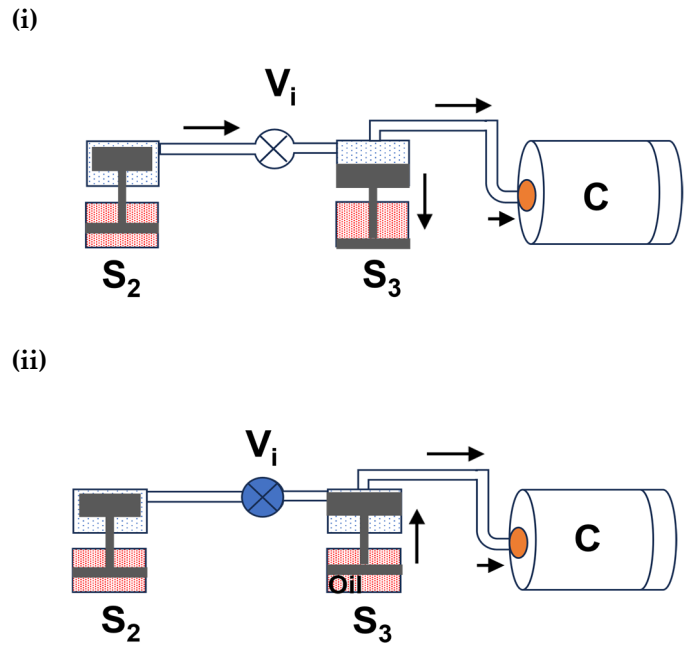

**(ii)**

**Figure: 3.** The block diagram of the third stage gas compression process: **(i)** Gas flow into the $S_3$ cavity and chamber (C) **(ii)** $S_3$ piston compressing the gas into chamber (C) up to 1.8 GPa (18000 bar).

## 3. HP-HTS technique at UNIPRESS

HP-HTS facility designed at UNIPRESS is capable of producing a pressure up to 1.8 GPa and temperature up to 1700°C. One can use a one- or multi-zone furnace to create different temperatures, especially for a single crystal growth process [9, 8, 11, 14]. Our current system is based on a three-stage oil-gas compressor to create high pressure as discussed above. More advanced designs can produce pressure up to 3 GPa, which requires three pistons and a small diameter of the pressure chamber (C). Our HP-HTS technique is vital in synthesizing high-quality materials [9, 8, 11, 14]. With the capability to generate high gas pressure, this system has multi-zone furnaces, real-time temperature measurement, and adaptability for various sample types positions inside the pressure chamber. It is a versatile tool for growing single and polycrystalline samples across diverse material categories [8]. The block diagram of our HP-HTS system is depicted in Figure 4. The system has four main components: a high-pressure chamber, compressor, sample holder with furnace, and controller unit and monitor, which work cohesively to ensure precise and controlled pressurization, leak-free operation, and real-time temperature and pressure monitoring. These features empower high-pressure synthesis and sintering research, providing a platform for advancements in superconductivity research and applications.

HP-HTS system which is currently used for the growth of FBS, is based on three pistons, as depicted in Figure 4. These pistons are connected to each other and also to the pressure chamber. The first stage ($S_1$) generates the pressure to the second stage ($S_2$), the third stage ($S_3$), and the pressure chamber (C). In the next step, the first stage is disconnected and the second-stage piston creates the pressure to the third stage ($S_3$) and pressure chamber. Finally, the third stage ($S_3$) starts to work and generate the maximum pressure inside the chamber up to 1.8 GPa. More details about each part are given below:

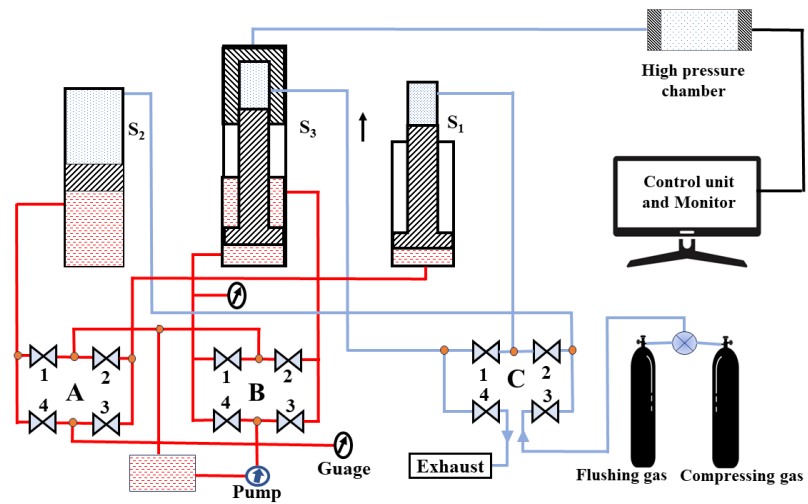

**Figure 4**. The block diagram of HP-HTS technique, presented at our institute, consisting of a three-stage oil-based compressor, high-pressure chamber (C), and a control unit monitor.

*i) High-Pressure chamber:* The high-pressure chamber is a central component, depicted in Figure 5, and is constructed from robust steel to withstand extremely high pressures. Its intricate design ensures that this camber remains impervious to leaks even under high-pressure conditions. To regulate temperature efficiently, the outer jacket of the chamber is integrated with a water cooling system. One side of the chamber is precisely connected to a capillary using a large-diameter screw, supported with O-rings and additional seals, ensuring a completely leak-free chamber. This capillary tube performs a dual purpose, functioning as both the inlet and outlet for gas, which is connected to the compressor and gas bottle. On the opposite side, a sample holder with a high-quality O-ring and metal seal, facilitates the insertion of samples into the pressure chamber, as shown in Figure 5. A solder plug is thoughtfully employed to route wires, and safely eliminate leakage risk. These wires are connected to the control unit and monitor, forming a reliable and secure connection for the entire system. The chamber's adaptability for single, double, or triple zone furnaces is a noteworthy feature, enhancing its utility for growing high-quality single crystals and facilitating a wide range of experiments in materials science.

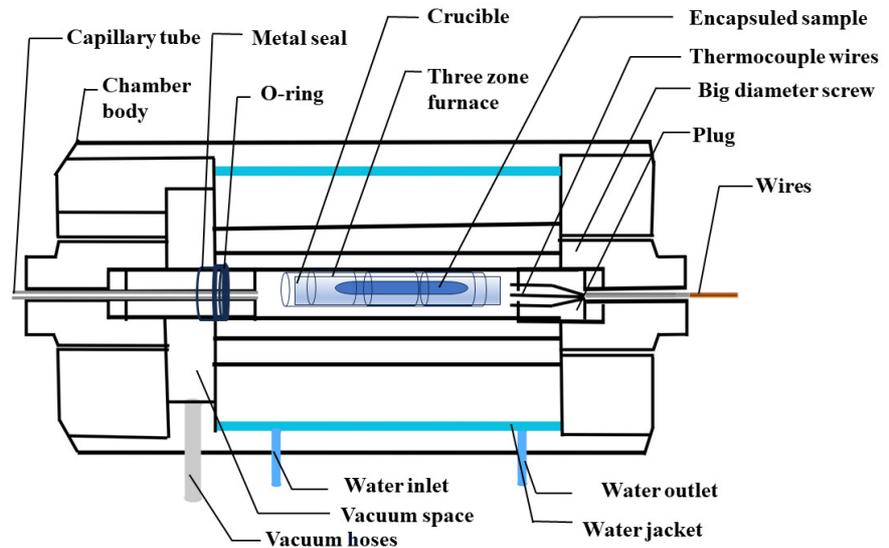

**Figure: 5.** Schematic diagram of high pressure chamber.

*ii) Compressor*: The compressor is an oil-based system with a three-stage piston configuration with pump, as shown in Figure 6. The piston moves upward to create the gas pressure. Generally, this pump is connected to all the stages and moves the piston upward to create high pressure. The systematic block diagram is shown in Figure 6, and the list of maximum pressure for different stages is mentioned in Table 1. In the first stage ($S_1$), it can generate a pressure up to 800 bar. The second piston ($S_2$) can achieve pressures of up to 4000 bar and in the final stage ($S_3$), it can reach an impressive pressure of up to 1.8 GPa (18000 bar). This three-stage piston setup allows for the precise control and adjustment of pressure from low to very high range, which is one of the notable strengths of our compressor. The presence of 12 key valves is the most intricate and critical aspect of the compressor, as shown in Figure 6, which plays a pivotal role during the gas compression process which ensures precise and controlled pressurization. The basic principle of the pump is already explained above.

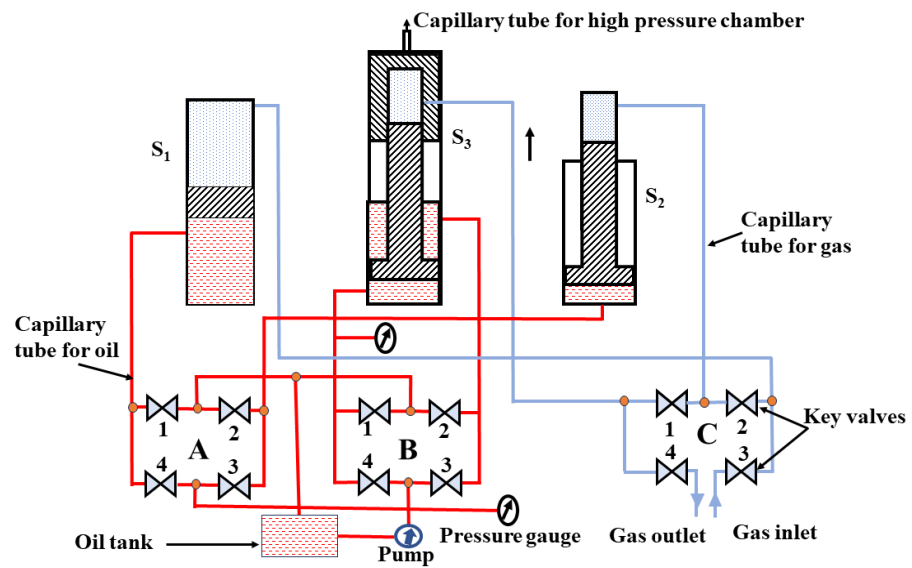

**Figure 6:** Schematic diagram of a compressor based on the oil pump with three-stage pistons.

**Table 1.** List of the maximum pressure created by the different stage of a compressor.

| Stages of Piston | Maximum Pressure |
|---|---|
| First stage ($S_1$) | 800 bar / 80 MPa |
| Second stage ($S_2$) | 4000 bar / 400 MPa |
| Third stage ($S_3$) | 18000 bar / 1.8 GPa |

*iii) Sample holder with furnace:* This component contains a wide setup, consisting of essential components like the sample container, furnace, thermocouples, and pressure gauge. A systematic block diagram is shown in Figure 7. The sample container, designed either in a boat or cylindrical form with a secure cap, acts as the vessel for keeping the samples inside the high-pressure chamber. The furnace is equipped with a Kanthal (FeCrAl alloy) heater wire capable of reaching temperature up to 1300°C. When higher temperatures are needed, molybdenum (Mo) and tungsten wires can be utilized through which the maximum temperature can reach up to approximately 1700°C within this high-pressure chamber [3], as listed in Table 2. Furthermore, either a single or multi-zone furnace can be used as shown in Figure 7 to create the high temperature. These thermocouples and heaters are interfaced to the computer through software.

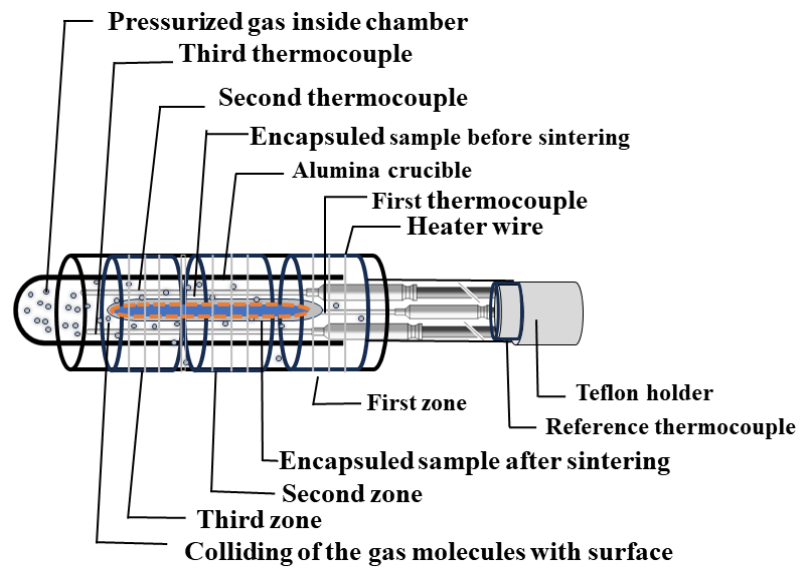

**Figur: 7.** Block diagram of sample holder with furnace.

For accurate temperature monitoring, three thermocouples are employed, in addition to a reference thermocouple that provides precise temperature readings across all zones. We can use various kinds of thermocouples, as mentioned in Table 3. These are typically K-type thermocouples made of Chromel and Nickel-Aluminium, and are suitable for temperatures up to 1100°C. However, various thermocouples can be selected based on specific requirements [15, 3]. The actual pressure inside the chamber is regularly monitored using pressure gauges. These vital components, including the thermocouples, heaters, and gauges, are seamlessly connected to the controller unit and monitor, ensuring precise control and monitoring of the high-pressure chamber's conditions. Modifying this chamber according to specific material growth conditions adds to its versatility and suitability for diverse research and experimentation needs in materials science.

Table 2. Heater wires for the furnace with a range of maximum temperature.

| Wires Materials | Description |
| --- | --- |
| Tungsten (W) | High resistive metal up to 2600 °C |
| Molybdenum (Mo) | High resistive metal up to 2200 °C |
| Kanthal (FeCrAl alloy) | High resistivity up to 1400°C |
| Nifethal (NiFe alloy) | Low resistive up to 600°C |

Table 3. Types of thermocouples that can be used for HP-HTS technique.

| Thermocouples | Temperature Range (°C) | Composition |
|---|---|---|
| **B-Type** | 200 to 1700 | (+) Platinum- 6% Rhodium <br> (−) Platinum- 30% Rhodium |
| **K-type** | 0 to 1100 | (+) Nickel-Chromium <br> (−) Nickel-Aluminum |
| **T-Type** | -200 to 370 | (+) Copper <br> (−) Constantan |

*iv) Controller unit and monitor:* The final component of HP-HTS comprises a temperature controller and a pressure gauge, both of which can be connected to a computer for enhanced functionality. This capability allows for real-time monitoring of temperature and pressure within the system. The temperature controller facilitates monitoring all four thermocouples, each positioned at different locations within the system. One of these thermocouples also serves as a reference temperature point, enabling precise temperature measurements near the sample. This feature provides valuable insights into the actual temperature conditions during experiments. Through the interface with the computer, the pressure gauge provides real-time data on the pressure inside the chamber. This allows for continuous monitoring and control, ensuring that pressure conditions are maintained as required for specific experiments.

## 4. Current results using this HP-HTS facility

Currently, we are applying this HP-HTS technique for FBS [16, 17] which has the highest $T_c$ of 58 K, high upper critical field ($H_{c2}$) of 100 T, and high critical current density ($J_c$) of $10^7$-$10^8$ A/cm$^2$ at 5 K [18]. These properties make them a strong contender for the practical applications [19]. Many compounds belonging to this high $T_c$ superconductor can be categorized into six families in which 1111 (*RE*FeAs(O,F), *RE* = La, Ca, Pr, Gd) as a doped family and 1144 (*AeA*Fe$_4$As$_4$; *Ae* = Ca; *A* = K) as a stoichiometric family provide the highest $T_c$ of 58 K [20] and 36 K [21] respectively, for FBS. Hence, our current focus is the growth process of 1111 and 1144 families by using the HP-HTS technique [14]. Before using the high-pressure technique for these complicated families, HP-HTS technique was applied for the simplest FBS *i.e.* 11 (FeSe) family. Tellurium (Te) doping at Se-sites *i.e.* Fe(Se,Te) provides the highest transition temperature $T_c$ of 15 K for FeSe$_{0.5}$Te$_{0.5}$ [22]. First, we have synthesized the high-quality Fe(Se,Te) sample by solid-state reaction method at ambient pressure (0 GPa). The selected composition FeSe$_{0.5}$Te$_{0.5}$ is prepared at 600°C for 21 h in the first step, and in the second stage, the sample is grounded and heated again at 600°C for 4 hours, as more detail discussed in our previous studies [22, 11]. This FeSe$_{0.5}$Te$_{0.5}$ sample shows the transition temperature up to 15 K by the convenient synthesis method at ambient pressure, as reported by the previous report [22] and depicted in Figure 8(a).

To understand the high-pressure growth effect, Fe(Se,Te) bulks are prepared in a very broad pressure range from 0 GPa to 1 GPa at 600°C for 1 h and 11 hours, as reported in our previous study [11]. Also, these samples were prepared by *in-situ* and *ex-situ* processes where samples were sealed in a Ta-tube or placed in an open Ta-tube. These various conditions were used to optimize the best growth conditions so that the high-quality sample can be produced with high superconducting properties. Interestingly, the optimized conditions were obtained as 600°C, 1 h, and 500 MPa where samples show the highest superconducting properties [11]. The comparative graph is shown in Figure 8(a). Our studies also confirm that grain connectivity can be improved when the samples are sealed into a Ta-tube under an argon gas atmosphere through ARC melter, whereas the samples placed in an open Ta-tube have a pure superconducting phase but the grain connectivity was poor due to the high gas pressure passing through the micro or nanopores. Fe(Se,Te) bulks prepared by the optimized conditions are depicted in Figure 8(a), which shows the high superconducting properties ($T_c$ = 17 K) compared to the samples prepared by the ambient pressure method ($T_c$ = 15 K). These optimized study processes confirm that the high gas pressure technique can be an effective method to enhance the superconducting properties and the synthesis process can be completed in a very short reaction time under the optimized growth pressure [11].

After having a good experience with this 11 family of FBS, we have started to work with CaKFe$_4$As$_4$ (1144) which is a stoichiometric compound of FBS [14, 23]. On the basis of the optimization of the 11 (FeSe$_{0.5}$Te$_{0.5}$) family, we have prepared 1144 samples under the optimized conditions (500 MPa, 1 hour) at 500°C. Interestingly, CaKFe$_4$As$_4$ prepared by HP-HTS has enhanced the superconducting transition by ~2 K with improved sample quality compared to the 1144 bulks prepared under the conventional synthesis method [14] at ambient pressure. The resistivity measurements of these samples are depicted in Figure 8(b) which clearly enhanced the $T_c$ value with a sharp transition. It suggests that 1144 bulk prepared by HP-HTS are homogeneous and well-connected grain boundaries. In a similar way, we are also applying this HP-HTS to other members of 1144 and 1111 families.

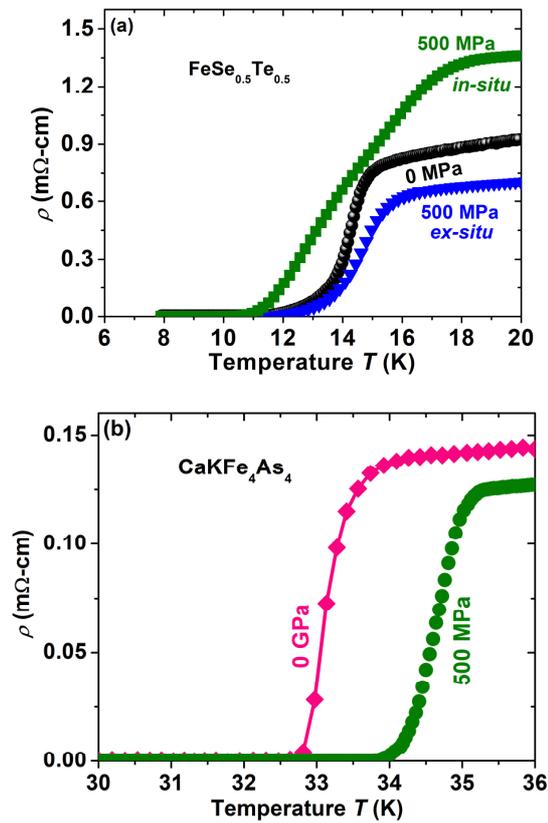

**Figure: 8. (a)** The temperature dependence of the resistivity ($\rho$) for **(a)** FeSe$_{0.5}$Te$_{0.5}$ [11] **(b)** CaKFe$_4$As$_4$ (1144) bulk [14] under HP-HTS, presented at our institute.

## 5. Conclusions

High gas pressure technique can be a unique way to improve the sample quality, sample size, and the material properties. HP-HTS facility available at our institute provides the larger sample space and high growth temperature and pressure. We can also create different gas atmosphere according to our requirements. The current application of this HP-HTS technique is going for high-temperature iron-based superconductors. Interestingly, the observed results depict the enhancement of superconducting properties and also the improved sample quality. Our studies suggest that the high-pressure synthesis works well for the high $T_c$ material and can be useful for other kinds of materials to improve their properties.

**Author Contributions:** Conceptualization, Supervision and Formal analysis, S.J.S.; methodology, S.J.S., M.A. and M.M.; data collection, M.M. and M.A.; High-pressure experiments, M.A., M.M., A.M. and T.C.; investigation and writing—original draft preparation, S.J.S., M.A. M.M.; writing—review and editing, S.J.S., M.A. and M.M.; Comments and Suggestions, M.A., M.M., A.M., S.J.S.. All authors have read and agreed to the published version of the manuscript.

**Funding:** This research was funded by National Science Centre (NCN), Poland, grant number "2021/42/E/ST5/00262" (SONATA-BIS 11). S.J.S. acknowledges financial support from National Science Centre (NCN), Poland through research Project number: 2021/42/E/ST5/00262.

## References

[1] M. H. Bocannegra-Bernal, "Review: Hot Isostatic Pressing (HIP) technology and its applications to metals and ceramics," *J. Mater. Sci.,* vol. 39, p. 6399, 2004.

[2] V. V. Brazhkin, "High-pressure synthesized materials: Treasures and hints," *High Pressure Research,* vol. 27, p. 333, 2007.

[3] M. Koizumi, "Hot isostatic pressing- Theory and applications," in *Third International Conference*, Osaka Japan, 1991.

[4] M. Fujioka, S. J. Denholme, M. Tanaka, H. Takeya, T. Yamaguchi, Y. Takano, "The effect of exceptionally high fluorine doping on the anisotropy of single crystal single crystalline $SmFeAsO_{1-x}F_x$," *Appl. Phys. Lett.,* vol. 105, p. 102602, 2014.

[5] H. V. Atkinso, S. Davies, "Fundamental Aspects of Hot Isostatic Pressing: An Overview," *Metall. Mater. Trans. A: Phys.,* vol. 31, p. 2983, 2000.

[6] J. Karpinski et al., "Single crystals of $LnFeAsO_{1-x}F_x$ (Ln = La, Pr, Nd, Sm, Gd) and $Ba_{1-x}Rb_xFe_2As_2$," *Physica C,* vol. 469, p. 370, 2009.

[7] L. N. Sang et al., "Pressure effects on iron-based superconductor families: Superconductivity, flux pinning and vortex dynamics," *Materials Today Physics,* vol. 19, p. 100414, 2021.

[8] J. Karpinski, H. Schwer, I. Mangelschots, K. Conder, A. Morawski, T. Lada, A. Paszewin, "Single crystals of $Hg_{1-x}Pb_xBa_2Ca_{n-1}Cu_nO_{2n+2+\delta}$ and infinite-layer $CaCuO_2$. synthesis at gas pressure 10 kbar, properties and structure," *Physica C,* vol. 234, pp. 10-18, 1994.

[9] A. Morawski, T. Lada, A. Paszewin and K. Przybylski, "High gas pressure for HTS single crystals and thin layer technology," *Supercond. Sci. Technol,* vol. 11, pp. 193-199, 1998.

[10] R. Matsumoto et al., "High-Pressure Synthesis of Superconducting $Sn_3S_4$ Using a Diamond," *Inorg. Chem,* vol. 61, p. 4476–4483, 2022.

[11] M. Azam, M. Manasa, T. Zajarniuk et al., "High-Pressure Synthesis and the Enhancement of the Superconducting Properties of $FeSe_{0.5}Te_{0.5}$," *Materials,* vol. 16, p. 5358, 2023.

[12] O. Tkachenko, A. Morawski, A. J. Zaleski et al., "Synthesis, Crystal Growth and Epitaxial Layers Growth of $FeSe_{0,88}$ Superconductor and Other Poison Materials by Use of High Gas Pressure Trap System," *J Supercond. Nov. Magn.,* vol. 22, p. 599–602, 2009.

[13] John B. West, "Robert Boyle's landmark book of 1660 with the first experiments," *J Appl Physiol,* vol. 98, pp. 31-39, 2005.

[14] Manasa, M.; Azam, M.; Zajarniuk, T.; Diduszko, R.; Cetner, T.; Morawski, A.; Wiśniewski, A.; Singh, S. J., "Enhancement of Superconducting Properties of Polycrystalline $CaKFe_4As_4$ by High-Pressure Growth," *Submitted to Jounal (Under review)*, 2023.

[15] N. L. Loh and K.Y. Sia, "An overview of hot isostatic pressing," *Journal of Materials Processing Technology,* vol. 30, pp. 45-65, 1992.

[16] S. J. Singh and M. Sturza, "Bulk and Single Crystal Growth Progress of Iron-Based Superconductors (FBS): 1111 and 1144," *Crystals,* vol. 12, p. 20, 2022.

[17] K. Iida, J. Hänisch, A. Yamamoto, "Grain boundary characteristics of Fe-based superconductors," *Supercond. Sci. Technol.,* vol. 33, p. 043001, 2020.